\documentclass[prd,twocolumn,a4paper]{revtex4}

\newcommand \p{\partial}
\newcommand \br{\nonumber \\ &&}

\begin{document}

\title{Hyperbolicity of second-order in space systems of evolution equations}

\author{Carsten Gundlach}
\affiliation{School of Mathematics, University of Southampton,
         Southampton SO17 1BJ, UK}
\author{Jos\'e M. Mart\'\i n-Garc\'\i a}
\affiliation{Instituto de Estructura de la Materia, Centro de F\'\i sica
Miguel A. Catal\'an, C.S.I.C., Serrano 123, 28006 Madrid, Spain}

\date{June 2005}


\begin{abstract}

A possible definition of strong/symmetric hyperbolicity for a
second-order system of evolution equations is that it admits a
reduction to first order which is strongly/symmetric hyperbolic. We
investigate the general system that admits a reduction to first order
and give necessary and sufficient criteria for strong/symmetric
hyperbolicity of the reduction in terms of the principal part of the
original second-order system. An alternative definition of strong
hyperbolicity is based on the existence of a complete set of
characteristic variables, and an alternative definition of symmetric
hyperbolicity is based on the existence of a conserved (up to lower
order terms) energy. Both these definitions are made without any
explicit reduction. Finally, strong hyperbolicity can be defined
through a pseudo-differential reduction to first order. We prove that
both definitions of symmetric hyperbolicity are equivalent and that
all three definitions of strong hyperbolicity are equivalent (in three
space dimensions). We show how to impose maximally dissipative
boundary conditions on any symmetric hyperbolic second order
system. We prove that if the second-order system is strongly
hyperbolic, any closed constraint evolution system associated with it
is also strongly hyperbolic, and that the characteristic variables of
the constraint system are derivatives of a subset of the
characteristic variables of the main system, with the same speeds.

\end{abstract}


\maketitle

\tableofcontents


\section{Introduction}

Research in numerical relativity has recently focused on 
obtaining a well-posed continuum initial-boundary value problem 
as a starting point for numerical time evolutions of systems such as a
black-hole binary. Well-posedness of an initial-boundary value problem
implies that an estimate 
\begin{equation}
\label{estimate}
||\delta u(\cdot, t)|| \le F(t) \left(||\delta u(\cdot, 0)||+\int_0^t ||\delta
  g(\cdot, \tau)||\,d\tau\right) 
\end{equation}
exists, where $u(x,t)$ is the solution, $u(x,0)$ the initial data,
$g(x,t)$ appropriate free boundary data, $\delta$ denotes a linear
perturbation and $||\cdot||$ stands for appropriate norms (which may
involve spatial derivatives), and where $F(t)$ is independent of the
initial and boundary data. This means that the solution depends
continuously on the initial and boundary data.  Hyperbolicity is a
property of the evolution equations that can be used as an algebraic
criterion for well-posedness. We briefly review several notions of
hyperbolicity.

Consider a system of quasilinear evolution equations that is first
order in both space and time, or
\begin{equation}
\label{firstorder}
\dot u = P^i(u) u_{,i} + S(u),
\end{equation}
where $u$ is a vector of variables and $P^i$ are square matrices. 

{\noindent{\bf Definition 1:} {\em
The system (\ref{firstorder}) is called weakly hyperbolic if the
matrix $P^n\equiv n_iP^i$ has real eigenvalues for any unit vector
$n_i$.}

{\noindent{\bf Definition 2:} {\em The system (\ref{firstorder}) is
called strongly hyperbolic if $P^n$ is diagonalisable with real
eigenvalues for any $n_i$, and the matrix $T_n$ that diagonalises it and
its inverse $T_n^{-1}$ depend smoothly on $n_i$.}

{\noindent{\bf Definition 3:} {\em 
The system (\ref{firstorder}) is called symmetric hyperbolic if there
exists a Hermitian, positive definite matrix $H$ such that $HP^n$ is
Hermitian for any direction $n_i$ and where $H$ does not depend on
$n_i$.}

The following properties of strongly and symmetric hyperbolic systems
give a more practical meaning to the definitions, and we shall use
them later to {\em define} strong and symmetric hyperbolicity for
second-order systems. The key concept for strong hyperbolicity is

\noindent{\bf Definition 4:} {\em A characteristic variable with speed
$-\lambda$ in the $n_i$ direction is a linear combination $\bf u$ of
the variables $u$ that obeys
\begin{equation}
\label{Udef}
\partial_t {\bf u}= \lambda\partial_n {\bf u} + \dots,
\end{equation}
where $n_i$ is normalised with respect to some metric,
$\p_n\equiv n^i\p_i$, and the dots denote derivatives transverse to
$n_i$ with respect to the same metric, and lower order terms.}

If we write {\bf u} as ${\bf u}=\bar u^\dagger u$, where $\bar u$ is a
constant vector of coefficients, then
\begin{eqnarray}
\partial_t(\bar u^\dagger u)&=&\bar u^\dagger P^n \partial_n u + \dots 
=\lambda \p_n (\bar u^\dagger u) + \dots
\end{eqnarray}
if and only if $\bar u^\dagger$ is a left eigenvector of $P^n$ or
equivalently if $\bar u$ is an eigenvector of
$P^{n\dagger}$. Characteristic variables $\bf u$ of the first-order
reduction therefore correspond to left eigenvectors $\bar u^\dagger$
of $P^n$. This gives us

\noindent{\bf Lemma 1:} {\em A first-order system is strongly
hyperbolic if and only if it admits a complete set of characteristic
variables with real speeds that depend smoothly on $n_i$.}

The key concept for symmetric hyperbolicity is that of an energy:

\noindent{\bf Definition 5:} {\em An energy $\epsilon$ is a quadratic
form in $u$ that is positive definite in the sense that $\epsilon=0$
if and only if $u=0$, and which is conserved in the sense that there
exists a flux $\phi^i$ quadratic in $u$ such that}
\begin{equation}
\dot\epsilon = {\phi^i}_{,i}.
\end{equation}

With 
\begin{equation}
\epsilon\equiv u^\dagger Hu ,
\qquad \phi^i \equiv u^\dagger HP^iu,
\end{equation}
we have 

\noindent{\bf Lemma 2}: {\em A linear first-order system with
constant coefficients is symmetric hyperbolic if and only if it admits
an energy.}

For quasilinear systems this energy is conserved in the approximation
where $S(u)$ is neglected and $P^i(u)$ is approximated as
constant. (Physically, this corresponds to considering small
high-frequency perturbations $\delta u$.) When boundaries are present,
the time derivative of the energy can be estimated in terms of free
boundary data.

Strong hyperbolicity of a first-order system is necessary and
sufficient for a well-posed Cauchy problem. The Cauchy problem for a
merely weakly hyperbolic system is typically ill-posed in the presence
of lower-order terms. Symmetric hyperbolicity implies strong
hyperbolicity, and is therefore also sufficient for well-posedness of
the Cauchy problem. Furthermore, symmetric hyperbolicity can be used to prove
well-posedness of the initial-boundary value problem for a certain
class of boundary conditions called maximally dissipative \cite{GKO}.

Hyperbolicity for equations or systems of equations of higher than
first order is less well-established. A definition of weak hyperbolicity
exists for systems of arbitrary order, but as for first-order systems,
it does not guarantee well-posedness \cite{Beig}. Alternatively,
a quasilinear system that is second order in both space and time, or
\begin{equation}
P^{\mu\nu}(u,\partial u)u_{,\mu\nu}+ S(\partial u,u)=0,
\end{equation}
is called hyperbolic if $P^{\mu\nu}$ is a Lorentzian metric, that is
if the principal part of the system is that of a wave equation
\cite{Wald}. Christodoulou has recently generalized the idea
introducing the concept of regular hyperbolicity, with less strict
positivity requirements on the elliptic block of the principal part
\cite{reghyp}. Both can be used as criteria for
well-posedness of the Cauchy problem. The Einstein equations are
second order, but they fit these definitions of hyperbolicity only
when written in harmonic gauge, and so there are no standard
definitions of hyperbolicity immediately applicable to forms of the
Einstein equations commonly used in numerical relativity.

One possible approach to the well-posedness of a second-order system is
to reduce it to first order by introducing auxiliary variables, and to
define the second-order system to be strongly hyperbolic or symmetric
hyperbolic if the reduction is. An ad-hoc definition along those lines
has been used by Sarbach and co-authors \cite{SarbachBSSN,BeyerSarbach} to
prove well-posedness of the BSSN formulation of the Einstein
equations. We formalise this approach in Section~\ref{section:first}.

Independently, Nagy, Ortiz and Reula \cite{NOR}, following 
Kreiss and Ortiz \cite{KreissOrtiz} have used a pseudo-spectral
reduction to define strong hyperbolicity. This method does not appear
to generalise to symmetric hyperbolicity, intuitively because Fourier
transforms cannot be carried out on a domain with arbitrary boundary.
We briefly review this approach in Section~\ref{section:pseudo}.
By casting it in the notation of Section~\ref{section:first}, show
that the two definitions of strong hyperbolicity are equivalent. 

As a third alternative, Gundlach and Mart\'\i n-Garc\'\i a
\cite{bssn1,bssn2} define strong and symmetric hyperbolicity directly
from the second-order system, by focusing on the existence of
characteristic variables in strong hyperbolicity, and of an energy in
symmetric hyperbolicity. We review this approach in
Section~\ref{section:second}, and show that its definitions of both
strong and symmetric hyperbolicity are equivalent to those using a
first-order reduction. 

Outside the main line of this paper, we analyse in
Section~\ref{section:originalconstraints} the well-posedness of the
propagation of any constraints that the original second-order system
is subject to. In Section~\ref{section:symmpar} we apply our results
for symmetric hyperbolic systems to mixed symmetric
hyperbolic-parabolic systems. Section~\ref{section:conclusions}
summarises our results.


\section{The system}
\label{section:system}

In this short Section, we establish notation for the class of system
that we want to investigate, and clarify the relation between systems
that are second order in both time and space, only in space, or only
in space and that only in some of the variables. 

We begin systems that are first order in time. Formulations of the
Einstein equations based on the ADM formulation are naturally first
order in time and second order in space. Some cannot even be written
in second-order in time form (for example because they have an odd
number of variables). With a non-zero shift, the first-order form may
also be preferable for numerical simulations \cite{shiftedwave}. For
simplicity, we restrict attention to linear systems with constant
coefficients. These can be considered as the linearisation and frozen
coefficients approximation of a nonlinear system. 

The class of systems that we are interested in are not uniformly
second-order: some variables $u$ may appear in the evolution equations
without second spatial derivatives. Writing $u=(v,w)$, where $w$ are
those variables that appear only with first derivatives, we consider
therefore 
\begin{eqnarray}
\label{vdot0}
\dot v&=&A_1^{ij}v_{,ij}+A_1^iv_{,i}+A_1 v+A_2^i w_{,i}+A_2 w+a, \\
\label{wdot0}
\dot w&=&B_1^{ij}v_{,ij}+B_1^iv_{,i}+B_1 v+B_2^i w_{,i}+B_2 w+b.
\end{eqnarray}
Here $v$ and $w$ are column vectors (not necessarily of the same
length) of variables, the capital letters represent constant matrices
of the appropriate dimension, and $a$, $b$ are forcing functions.  We
assume that while at least one second derivative of every
variable $v$ appears in the equations, the number of variables $v$ has
been minimised. 

To reduce the system (\ref{vdot0},\ref{wdot0}) to first order, we define
the auxiliary variables $d_i\equiv v_{,i}$. By taking a spatial
derivative of (\ref{vdot0}), we find
\begin{equation}
\dot d_i=A_1^{jk}v_{,ijk}+A_1^jv_{,ij}+A_1 v_{,i}+A_2^jw_{,ij}+A_2
w_{,i} +a_{,i}.
\end{equation}
This is of a higher order than we started from, unless $A_1^{ij}$ and
$A_2^i$ both vanish. (An example of a second-order evolution
equation that cannot be reduced to first order is the heat equation
$\dot u=u''$.) We have \cite{HinderPhD}

{\noindent{\bf Lemma 3:} {\em 
The general second-order in space, first-order in time
linear system that can be reduced to first order by the introduction of
auxiliary variables is of the form}
\begin{eqnarray}
\label{vdot}
\dot v&=&\underline{A_1^iv_{,i}}+A_1 v+\underline{A_2 w}+a, \\
\label{wdot}
\dot w&=&\underline{B_1^{ij}v_{,ij}}+B_1^iv_{,i}+B_1
v+\underline{B_2^i w_{,i}}+B_2 w+b.
\end{eqnarray}

From now on we refer to this as ``the'' second-order system.
We have underlined the highest derivatives. Without loss of generality
we assume from now on that $B_1^{ij}$ is symmetric.

In order to understand how general the system (\ref{vdot}-\ref{wdot}) is,
it is interesting to convert it into second-order in both space and
time form. Taking a time derivative of (\ref{vdot}) and using
(\ref{wdot}) to replace $\dot w$, we obtain
\begin{equation}
\ddot{v} = A_2 B_1^{ij} v_{,ij} + A_1^{i} \dot{v}_{,i} + A_2 B_2^iw_{,i} 
+ A_2 B_2 w + ...
\end{equation}
where we have written out all second derivatives and all appearances of
$w$. We can eliminate the remaining appearances of $w$ and $w_{,i}$
in terms of $\dot{v}$ using (\ref{vdot}) if and only if the matrices
of the system obey
\begin{equation} \label{ranks}
{\rm rank}(A_2) 
= {\rm rank}\left(\begin{array}{c} A_2 \\ A_2 B_2^i\end{array}\right)
= {\rm rank}\left(\begin{array}{c} A_2 \\ A_2 B_2\end{array}\right).
\end{equation}
When $A_2$ is invertible, which in particular implies equal numbers of
$v$ and $w$ variables, these conditions are automatically obeyed.
On the other hand, any fully second-order system in a set of variables $v$
can be reduced to the form (\ref{vdot}-\ref{wdot}) by introducing
$\dot v\equiv w$. Therefore the class of first-order in time,
second-order in space systems (\ref{vdot}-\ref{wdot}) includes the
class of fully second-order systems, but is much bigger.


\section{First-order reduction method}
\label{section:first}


\subsection{Parameterised reduction}
\label{section:reduction}

In reducing (\ref{vdot}-\ref{wdot}) to first order by defining
$d_i\equiv v_{,i}$, we can write each occurrence of $v_{,i}$ also as
$d_i$, or a mixture of the two, and similarly we can write $v_{,ij}$ as
$d_{i,j}$ or $d_{j,i}$. To parameterise these ambiguities, we formally
add multiples of the auxiliary constraint
\begin{equation}
c_i\equiv d_i-v_{,i}=0
\end{equation} 
and its antisymmetrised derivative 
\begin{equation}
c_{ij}\equiv d_{[j,i]}=c_{[j,i]}=0
\end{equation}
to all three equations. We could not add $c_{(i,j)}$, or any higher
derivatives of the auxiliary constraints, without increasing the order
of the system. The general reduction to first order is therefore
\begin{eqnarray}
\label{vdot1}
\dot v&=&A_1^iv_{,i}+A_1 v+A_2 w+a \br
+A_3^{ij} c_{ij}+A_3^i c_i, \\
\label{wdot1}
\dot w&=&B_1^{ij}d_{i,j}+B_1^iv_{,i}+B_1 v+B_2^i w_{,i}+B_2 w+b \br
+B_3^{ij} c_{ij}+B_3^i c_i, \\
\label{ddot1}
\dot d_i&=&A_1^j d_{i,j}+A_1 v_{,i}+A_2 w_{,i} +a_{,i} \br
+{D_i}^k c_k+{D_i}^{jk} c_{jk}.
\end{eqnarray}
From now on, we shall refer to this system as ``the'' reduction.
We shall refer to the constant matrices $A_3^i$, $A_3^{ij}$, $B_3^i$,
$B_3^{ij}$, ${D_i}^j$ and ${D_i}^{jk}$ as the reduction parameters. 
Without loss of generality, we assume that $A_3^{ij}$, $B_3^{ij}$ and
${D_k}^{ij}$ are antisymmetric in $i$ and $j$. The terms
in the second-order system that become principal terms in the
reduction are the ones underlined in (\ref{vdot}-\ref{wdot}). We
shall call these the principal part of the second-order system. 


\subsection{Auxiliary constraint evolution}

The evolution of $v$ and $d_i$ implies an evolution of the auxiliary
constraints $c_i$. This auxiliary constraint system can be written in
first-order in space and time form by introducing $c_{ij}\equiv
c_{[j,i]}$ (as already defined above) 
as auxiliary variables. This results in
\begin{eqnarray}
\label{cidot}
\dot c_i&=&A_1^jc_{i,j}-A_3^jc_{j,i}  - A_3^{jk}c_{jk,i} \br
+ {D_i}^j c_j + {D_i}^{jk} c_{jk} , \\
\label{cijdot}
\dot c_{ij}&=&A_1^k c_{ij,k}-{D_{[i|}}^k c_{k,|j]}
-{D_{[i|}}^{kl} c_{kl,|j]},
\end{eqnarray}
As the right-hand side of this system of linear PDEs is homogeneous,
$c(x,0)=0$ implies $\dot c(x,0)=0$. Assuming that the coefficients are
constant, as we do in this paper, on taking a Fourier transform in
$x^i$ we obtain a separate ODE for each wavenumber $\omega^i$, and it
follows that $c(x,t)=0$ is the {\em unique} solution with $c(x,0)=0$.}

We have shown that if the auxiliary constraints are zero initially,
they remain zero at all times.  This allows us to prove well-posedness
for the reduction to first order, and then restrict it to the subset
of solutions that obey the auxiliary constraints in order to infer
well-posedness of the original second-order system.

Note that in order to make this argument, well-posedness of the
auxiliary constraint system is not required, as we require only
existence and uniqueness of the zero solution, not estimates of any
non-zero solution.


\subsection{Definition of hyperbolicity}

We now focus on the principal part of the first-order reduction, 
\begin{equation}
\label{pp}
\partial_t u\simeq P^i\partial_i u, 
\end{equation}
where here and in the following
$\simeq$ denotes equality up to lower-order terms and now $u$
stands for $(v,w,d_i)$.   
The degree of hyperbolicity of the first-order system depends
on the the reduction parameters. The appropriate definitions of
hyperbolicity are therefore the following: 

{\noindent{\bf Definition 1a}: {\em  
The second-order system (\ref{vdot}-\ref{wdot}) is defined to be weakly
hyperbolic if and only if it admits at least one reduction to first
order (\ref{vdot1}-\ref{ddot1}) that is weakly hyperbolic.}

{\noindent{\bf Definition 2a}: {\em The second-order system is defined
to be strongly hyperbolic if and only if it admits at least one
reduction to first order that is strongly hyperbolic.}

{\noindent{\bf Definition 3a}: {\em The second-order system is defined
to be symmetric hyperbolic if and only if it admits at least one
reduction to first order that is symmetric hyperbolic.}

In the remainder of this Section we derive necessary and sufficient
conditions for these definitions to hold, formulated directly in terms
of the principal part of the second-order system, without reference to
a reduction. 


\subsection{2+1 split}

We introduce a matrix notation in the groups of variables $v$,
$w$ and $d$. In this notation, (\ref{pp}) is
\begin{equation}
\label{big}
\partial_t
\left(
\begin{array}{c}
v \\ w \\ d_j
\end{array}
\right)
\simeq P^i_{} \partial_i 
\left(
\begin{array}{c}
v \\ w \\ d_k
\end{array}
\right) 
\end{equation}
where 
\begin{equation} \label{P1}
P^i_{} 
\equiv 
\left(
\begin{array}{ccc}
A_1^i-A_3^i & 0 & A_3^{ik} \\
B_1^i-B_3^i & B_2^i & B_1^{ik}+B_3^{ik} \\
A_1 {\delta_j}^i-{D_j}^i & A_2 {\delta_j}^i & A_1^i {\delta_j}^k
+{D_j}^{ik} 
\end{array}
\right).
\end{equation}
We expand the tensor indices $i$, $j$ and $k$ in (\ref{P1}) into their
components in the direction $n_i$ and transverse to it. For this
purpose we need a positive definite metric tensor $\gamma_{ij}$. This
can be chosen to be $\delta_{ij}$, or the physical 3-metric
for example in applications to general relativity. We
can then define the normal component of a tensor such as $d_n\equiv
n^id_i$ where $n^i\equiv \gamma^{ij}n_j$ and $n_i$ is normalised so that $n_i
n_j \gamma^{ij}\equiv 1$, and its transverse part such as
$({\delta_i}^j-n_in^j)d_j$, which we denote by $d_A$. For a reason
that will become apparent immediately, we also re-order the rows and
columns. We call the resulting matrix ${\cal P}$. It is related to
$P^n$ by a unitary transformation that depends on $n_i$, but it is not
the $n_i$ component of any vector of matrices. We have
\begin{equation}
\partial_t
\left(
\begin{array}{c}
w \\ d_n \\ v \\ d_A 
\end{array}
\right)
\simeq {\cal P}_{} \partial_n 
\left(
\begin{array}{c}
 w \\ d_n \\ v \\ d_B
\end{array}
\right) + \hbox{transverse derivatives},
\end{equation}
where 
\begin{equation} 
{\cal P}
\equiv 
\left(
\begin{array}{cccc}
B_2^n & B_1^{nn} & B_1^n-B_3^n & B_1^{nB}+B_3^{nB}   \\
A_2 & A_1^n & A_1 -{D_n}^n & {D_n}^{nB}  \\
0 & 0 & A_1^n-A_3^n & A_3^{nB}  \\
0 & 0 & -{D_A}^n & A_1^n {\delta_A}^B +{D_A}^{nB} \\
\end{array}
\right).
\end{equation}
We write this in shorthand form as 
\begin{equation} 
\label{calAdef}
{\cal P}_{} 
\equiv 
\left(
\begin{array}{cc}
{\cal A} & {\cal B} \\ 
0 & {\cal C} \\ 
\end{array}
\right),
\quad 
{\cal A}
\equiv 
\left(
\begin{array}{cc}
 B_2^n & B_1^{nn} \\ 
 A_2 & A_1^n 
\end{array}
\right).
\end{equation}
The eigenvalues of ${\cal P}$ are those of $P^n$, and one is
diagonalisable if and only if the other is. Therefore we can
investigate weak and strong hyperbolicity using ${\cal P}$.
The fact that ${\cal P}$ has a zero block for all choices of the
reduction parameters has several interesting consequences for its
eigenvalues and eigenvectors, which respectively determine the weak
and strong hyperbolicity properties of the first-order system.


\subsection{Eigenvalues of ${\cal P}$: weak hyperbolicity}

The first of these consequences is that the eigenvalues of $\cal P$
(and hence of $P^n$) are the union of the eigenvalues of ${\cal A}$
and $\cal C$, and are independent of $\cal B$.  We will show in
Section~\ref{section:strong} that we can choose the reduction
parameters so that $\cal C$ has real eigenvalues. This gives us

{\noindent{\bf Lemma 4:} {\em The second-order system is weakly
hyperbolic according to Definition 1a if and only if ${\cal A}$ has
real eigenvalues for all $n_i$.}

We can further analyse the eigenvalues of ${\cal A}$. If we replace $n_i$ by
$-n_i$, then $B_2^n$ and $A_1^n$ change signs, while $B_1^{nn}$ and
$A_2$ do not. This means that if $\lambda$ is an eigenvalue of ${\cal A}$ for
some $n_i$ then $-\lambda$ is an eigenvalue of ${\cal A}$ for $-n_i$. 
The eigenvalues must therefore either be $\lambda=O(n_i)$ where $O$ is
odd in $n_i$, or they are paired as $\lambda_\pm=O(n_i)\pm E(n_i)$ where
$O$ is odd and $E$ is even. As a consequence, if the dimension of ${\cal A}$
is odd, at least one eigenvalue must be of the form $\lambda=O(n_i)$,
and by continuity it must vanish for some $n_i$.


\subsection{Eigenvectors of ${\cal P}$: strong hyperbolicity}
\label{section:strong}

The reduction is strongly hyperbolic if and only if $P^n$ (or
equivalently ${\cal P}$) is diagonalisable. In
Appendix~\ref{appendix:diagonalisation} we show that a necessary
condition for ${\cal P}$ to be diagonalisable is that both ${\cal A}$
and $\cal C$ are diagonalisable. If the sets of eigenvalues of ${\cal
A}$ and $\cal C$ are disjoint, this is also sufficient. If they have
any eigenvalues in common, additional necessary criteria arise which
involve $\cal B$.

To state these conditions in a simple form, diagonalise ${\cal A}$ and
$\cal C$ simultaneously, so that
\begin{equation}
S^{-1}{\cal P}S=
\left(
\begin{array}{cc}
\Lambda_{\cal A} & \tilde {\cal B} \\ 
0 & \Lambda_{\cal C} \\ 
\end{array}
\right) 
\end{equation}
where $\Lambda_{\cal A}$ and $\Lambda_{\cal C}$ are diagonal matrices
with the eigenvalues of $A$ and $\cal C$ respectively in the diagonal.  
Let all repeated eigenvalues be grouped together in these matrices.
Then, for each eigenvalue common to ${\cal A}$ and $\cal C$,
the {\em corresponding block} of $\tilde {\cal B}$ must vanish if
${\cal P}$ is to be diagonalisable.

The block ${\cal A}$ does not contain any reduction parameters, and
so is determined by the original second-order system.  For a given
direction $n_i$, blocks $\cal B$ and $\cal C$ are determined completely by the
choice of reduction parameters, but there are not enough reduction
parameters to make this true for all directions $n_i$ at
once. (For example, as $B_1^{ij}$ is symmetric and $B_3^{ij}$ is
antisymmetric in $ij$, $B_1^{nB}+B_3^{nB}$ can be made to vanish for
any one direction $n_i$, but not for all directions.) 

We shall consider one choice of reduction parameters in
discussing strong hyperbolicity (here) and another one in discussing
symmetric hyperbolicity (in the next subsection). Both choices set
\begin{eqnarray}
A_3^{ij} &=& 0, \nonumber \\
A_3^i &=& A_1^i, \nonumber\\
B_3^i&=& B_1^i, \nonumber\\
{D_i}^j&=&A_1{\delta_i}^j,
\label{isolatev}
\end{eqnarray}
This partial choice has the effect of decoupling $v$ from the $w$ and
$d_i$. To discuss strong hyperbolicity in the case of three space
dimensions, we complete the choice of reduction parameters by
\begin{eqnarray}
B_3^{ij}&=& 0, \nonumber \\
\label{mychoice1}
{D_j}^{ik} &=& 
\delta_j{}^iA_1^k-\delta_j{}^kA_1^i+i\mu{\epsilon_j}^{ik},
\end{eqnarray}
where $\mu$ is a real constant, $\epsilon_{ijk}$ is the totally antisymmetric
tensor in three dimensions and $i=\sqrt{-1}$. This gives
\begin{equation} 
\label{myreduction2}
{\cal P}_{} 
=
\left(
\begin{array}{cccc}
B_2^n & B_1^{nn} & 0 & B_1^{nB}   \\
A_2 & A_1^n & 0 & A_1^B  \\
0 & 0 & 0 & 0  \\
0 & 0 & 0 & i\mu\epsilon_A{}^{nB} 
\end{array}
\right).
\end{equation}
The diagonal block $i\mu\epsilon_A{}^{nB}$ is diagonalisable with real
eigenvalues $\pm \mu$.  The complexification is unusual, but the term
multiplied by $i\mu$ is proportional to the auxiliary constraint
$c_{ij}$, and so has no influence on the original second-order
system. If we choose $\mu$ large enough, the eigenvalues of this block
are distinct from those of the complementary diagonal block
(containing $\cal A$ and a zero row and column), and we have found a
choice of reduction parameters that makes ${\cal P}$ diagonalisable if
${\cal A}$ is diagonalisable. The existence of this choice of
reduction parameters completes our proof of

{\noindent{\bf Theorem 1}: {\em The second-order system is strongly
hyperbolic according to Definition 2a if and only if ${\cal A}$ is
diagonalisable for all $n_i$ where the diagonalising matrix and its
inverse depend smoothly on $n_i$.}

Our proof of this theorem assumes three space dimensions, but we
suspect that the theorem holds in any number of space dimensions. 

Note that this criterion is based only on the coefficients on the
second-order system, without explicit reference to the reduction.
Note also that for the choice of reduction parameters we have used
here, the auxiliary constraint system is strongly hyperbolic, see
Appendix~\ref{appendix:auxiliary}.


\subsection{Symmetric hyperbolicity}
\label{section:redsymm}

According to Definition 3, the reduction is symmetric hyperbolic if
and only if there is a Hermitian matrix $H$ such that
\begin{equation}
\label{symm5}
(HP^n)^\dagger = HP^n
\end{equation} 
for all $n_i$, with $H$ independent of $n_i$ and positive
definite. Note that in the definition (\ref{symm5}) we cannot replace
$P^n$ by ${\cal P}$.

We again make the partial choice (\ref{isolatev}) of reduction
parameters. $B_3^{ij}$ and ${D_i}^{jk}$ will be determined in the
following in terms of $H$.  The resulting form of $P^i$ is [see
(\ref{big}) for the definition of $P^i$]
\begin{equation} 
\label{vseparate}
P^i_{} 
=
\left(
\begin{array}{ccc}
0 & 0 & 0 \\
0 & B_2^i & B_1^{ik}+B_3^{ik} \\
0 & A_2 {\delta_j}^i & A_1^i {\delta_j}^k
+{D_j}^{ik} 
\end{array}
\right).
\end{equation}
Clearly it is sufficient to find a symmetriser for the $(w,d_i)$
block. We parameterise $H$ as
\begin{equation}
\label{Hdef}
u^\dagger H u=(v^\dagger,w^\dagger,d_m^\dagger)
\left(
\begin{array}{ccc}
1 & 0 & 0 \\
0 & K & L^j \\
0 & L^{\dagger m} & M^{mj} 
\end{array}
\right)
\left(
\begin{array}{c}
v \\ w \\ d_j
\end{array}
\right)
\end{equation}
with $K$ and $M$ Hermitian and positive, $K^\dagger=K$, $K>0$, and
$M^{\dagger mj} = M^{jm}$, $M>0$. 

The nontrivial, $(w,d_i)$ block of $HP^i$ is
\begin{equation}\label{symhyp}
\left(
\begin{array}{cc}
KB_2^i+L^iA_2 & K(B_1^{ik}+B_3^{ik}) \\
& +L^kA_1^i+L^j{D_j}^{ik} \\
L^{\dagger m}B_2^i+M^{mi}A_2 & L^{\dagger m}(B_1^{ik}+B_3^{ik}) \\
& +M^{mk}A_1^i +M^{mj}{D_j}^{ik} 
\end{array}
\right).
\end{equation}
A necessary condition for this to
be Hermitian is for the matrix
\begin{equation}
\label{symhypn}
\left(
\begin{array}{cc}
KB_2^n+L^nA_2 & KB_1^{nn}+ L^nA_1^n \\
L^{\dagger n}B_2^n+M^{nn}A_2 & L^{\dagger n}B_1^{nn}+M^{nn}A_1^n
\end{array}
\right) 
\end{equation}
to be Hermitian for all $n_i$. 
This is just the condition that ${\cal A}$ admits a symmetriser, or
\begin{equation}
\label{Asymm}
{\cal H}{\cal A}=({\cal H}{\cal A})^\dagger, \quad \hbox{where} \quad
{\cal H}\equiv
\left(
\begin{array}{cc}
K & L^n \\
L^{\dagger n} & M^{nn} 
\end{array}
\right),
\end{equation}
for all $n_i$.  If the system is strongly hyperbolic, ${\cal A}$
always admits a symmetriser formed from its
eigenvectors. Nevertheless, (\ref{Asymm}) is a non-trivial condition
because ${\cal H}$ must form a part of $H$, so that its blocks $L^n$ and
$M^{nn}$ are given by $L^in_i$ and $M^{ij}n_in_j$, where $K$, $L^i$
and $M^{ij}$ do not depend on $n_i$. We now show that (\ref{Asymm})
actually implies that all of (\ref{symhyp}) is Hermitian for a
particular choice of reduction parameters $B_3^{ij}$ and
${D_i}^{jk}$. We do this for each block of (\ref{symhyp}) in turn.

From the top left block of (\ref{symhyp}) we have
\begin{equation}
\label{T}
(K B_2^i+L^iA_2)^\dagger = K B_2^i+L^iA_2 .
\end{equation} 
This equation does not contain any reduction parameters, and it is
clearly equivalent to the condition of $K B_2^n + L^n A_2$ being
Hermitian for all $n_i$, which is contained in the first diagonal
block of (\ref{Asymm}).  

The off-diagonal blocks of (\ref{symhyp}) give
\begin{equation}
\label{Tik}
(L^{\dagger k}B_2^i+M^{ki}A_2)^\dagger =
K(B_1^{ik}+B_3^{ik})+L^kA_1^i+L^j{D_j}^{ik}.
\end{equation}
If we denote this equation by $T^{ik}=0$, then its symmetric part
$T^{(ik)}=0$, or
\begin{equation}
\label{T(ik)}
B_2^{\dagger (i}L^{k)}+ A_2^\dagger M^{(ik)} = K B_1^{ik},
+ L^{(k} A_1^{i)},
\end{equation}
again does not contain any reduction parameters. Furthermore
$T^{(ik)}=0$ if and only if $T^{ik}n_in_k=0$ for all $n_i$ (this is a
property of all totally symmetric tensors), and this is precisely the
condition contained in the off-diagonal terms of (\ref{Asymm}). The
antisymmetric part $T^{[ik]}=0$, or
\begin{equation}
\label{T[ik]}
B_2^{\dagger [i}L^{k]}+ A_2^\dagger M^{[ik]} = K B_3^{ik} +
L^j {D_j}^{ik},
\end{equation}
can be solved for $B_3^{ij}$ because $K$ is by assumption invertible.

Finally, the bottom right block of (\ref{symhyp}) gives
\begin{eqnarray} \nonumber
\label{Tmik}
L^{\dagger m}(B_1^{ik}+B_3^{ik})+M^{mk}A_1^i +M^{mj}{D_j}^{ik} \\
=\left[L^{\dagger k}(B_1^{im}+B_3^{im})+M^{km}A_1^i +M^{kj}{D_j}^{im}
\right]^\dagger.
\end{eqnarray}
If we write this as $T^{mik}=0$, the totally symmetric part
$T^{(mik)}=0$, or
\begin{eqnarray} 
\label{T(mik)}
L^{\dagger (m} B_1^{ik)}+M^{(mk}A_1^{i)}  
=\left[L^{\dagger (k} B_1^{im)}+M^{(km}A_1^{i)} 
\right]^\dagger.
\end{eqnarray}
does not contain any reduction parameters. It is
equivalent to $T^{mik}n_m n_i n_k=0$ for all $n_i$, and so vanishes
because of the last diagonal block of (\ref{Asymm}).
After solving (\ref{T[ik]}) for ${B_3}^{ij}$
we write (\ref{Tmik}) as
\begin{equation} \label{UMD}
U^{mik}+\bar{M}^{mj} {D_j}^{ik} =
U^{\dagger kim}+\left(\bar{M}^{kj} {D_j}^{im}\right)^\dagger,
\end{equation}
where we have defined 
\begin{eqnarray}
U^{mik} &\equiv& L^{\dagger m}B_1^{ik}+M^{mk}A_1^i \nonumber \\
&& + L^{\dagger m}K^{-1}\left(B_2^{\dagger[i}L^{k]}+A_2^\dagger M^{[ik]}
\right), \\
\bar{M}^{ij} &\equiv& M^{ij} -L^{\dagger i} K^{-1} L^j.
\end{eqnarray}
$\bar M$ is Hermitian ($\bar{M}^{\dagger ij} = \bar{M}^{ji}$) and
positive definite
[because $H$ is positive definite in particular when restricted
to vectors $(v,w,d_i)=(0,-K^{-1}L^kd_k,d_i$)], 
and hence invertible. We define 
\begin{equation}
\label{Xmikdef}
X^{mik}\equiv U^{\dagger kim}-U^{mik}, \qquad 
D^{mik}\equiv \bar{M}^{mj} {D_j}^{ik},
\end{equation}
and so
\begin{equation}
\label{Xmikantiherm}
X^{mik}=-X^{\dagger kim},
\end{equation}
while $T^{(mik)}=0$ is equivalent to
\begin{equation}
\label{Xmiksymm}
X^{(mik)}=0.
\end{equation}
(\ref{UMD}) becomes
\begin{equation}
D^{mik}-D^{\dagger kim}=X^{mik}, 
\end{equation}
which has the general solution
\begin{eqnarray}
\label{Dmiksolution}
6 D^{mik} &=& X^{kmi}+2X^{kim}+3 X^{ikm} \br
+4X^{imk}+5X^{mik} + Y^{mik},
\end{eqnarray}
using both (\ref{Xmikantiherm}) and (\ref{Xmiksymm}). The object
$Y^{mik}$ must obey
\begin{equation}
\label{Ymiksymm}
Y^{mik}=Y^{[mik]}, \qquad Y^{mik}=-Y^{\dagger mik},
\end{equation}
but is otherwise arbitrary. It parameterises the part of the
${D_j}^{ik}$ that is not determined by $H$, and can be set to zero. 

We have shown that a necessary and sufficient condition for $P^i$ to
admit a symmetriser $H$ is for ${\cal A}$ to admit a symmetriser
${\cal H}$ that depends on $n_i$ in the particular way given in
(\ref{Asymm}).  Therefore we have

{\noindent{\bf Theorem 2:} {\em A necessary and sufficient condition for
the second-order system  to by symmetric hyperbolic according to
Definition 3a is that (\ref{Asymm}) holds for all $n_i$ for some
$H>0$ parameterised by (\ref{Hdef}).}

Note that positive definiteness of $\cal H$ does not imply that of
$H$, because positive definiteness of $M^{nn}$ for all $n_i$ does not
imply positive definiteness of $M$. The difference can be expressed in
standard terminology as follows: a double quadratic form
${M^{ij}}_{AB}$ is rank-1 positive if and only if
${M^{ij}}_{AB}n_in_jm^Am^B >0$ for all $n_i$ and $m^A$; it is rank-2
positive if and only if ${M^{ij}}_{AB}n_i^A n_j^B >0$ for all
$n_i^A$. Rank-2 positivity implies rank-1 positivity, but they are not
equivalent when the indices $i,j,A,B$ belong to spaces of dimension 3
or larger, as shown in the example given in \cite{Ball}.  This
suggests it may be useful to introduce an intermediate concept of
hyperbolicity based on positivity of ${\cal H}$ for all $n_i$, rather
than positivity of $H$. This already guarantees that ${\cal A}$ is
diagonalisable, so that the system is strongly hyperbolic. The 
imposition of rank-1 positivity is also the key ingredient of the
definition of ``regular hyperbolicity'' by Christodoulou \cite{reghyp} for
second order in both space and time systems, which is also known to
yield well-posed problems. We believe there is a connection between
regular hyperbolicity and our condition ${\cal H}>0$, but have not
been able to show it.


\section{Direct second-order method}
\label{section:second}


\subsection{Strong hyperbolicity}

Following \cite{bssn1}, we now elevate Lemma 1 to a definition of
strong hyperbolicity for second-order in space, first-order in time
systems, with the only difference that the ${\bf u}$ become linear
combinations of $v_{,i}$ and $w$:

\noindent{\bf Definition 2b:} {\em The system (\ref{vdot}-\ref{wdot})
is called strongly hyperbolic if and only if is there is a set of
characteristic variables $\bf u$ linear in $w$ and $v_{,i}$ obeying
(\ref{Udef}) with real speeds $\lambda$ and where the matrix relating
$u$ and $\bf u$ and its inverse depend smoothly on $n_i$.}

This definition has two interesting consequences. The first is that
$v_{,A}$ can always be considered as a zero speed variable in the
direction $n_i$ because
\begin{equation}
\partial_t(v_{,A})=\partial_A(\partial_t v)
\end{equation}
so that the right-hand side can be considered as a sum of transverse
derivatives only. (In fact arbitrary linear combinations of the
$v_{,A}$ can be given arbitrary speeds.)

The second consequence is that nontrivial characteristic variables are
unique only up to adding transverse derivatives. If ${\bf u}$ is a
characteristic variable with speed $-\lambda$ then, for any
vector $X^A$ (made from the $u$)
\begin{eqnarray}
&& \partial_t({\bf u}+\partial_A X^A) \nonumber \\
&\simeq&\partial_t {\bf u}+\partial_A (\partial_t
  X^A)  \nonumber \\
&\simeq&\lambda \partial_n {\bf u} + \hbox{transv. deriv.}  \nonumber \\
&\simeq&\lambda \partial_n {\bf u} + \lambda \partial_A (\partial_n
  X^A) 
+ \hbox{transv. deriv.}
  \nonumber \\
&\simeq&\lambda \partial_n({\bf u}+\partial_A X^A)+ \hbox{transv. deriv.}
\end{eqnarray}
and so ${\bf u}+\partial_A X^A$ is also a characteristic variable with
speed $-\lambda$.  Such calculations rely on commuting partial
derivatives to interpret $\partial_A\partial_n\dots $ either as a
normal or a transverse derivative, depending on the situation.

The second-order system has no reduction parameters. However, for the
purpose of comparing the second-order approach with the reduction
approach, we can translate the second-order approach into the {\it
notation} of a first-order reduction. We account for the fact that
$v_{,i}$ and $d_i$, and $v_{,ij}$ and $v_{,ji}$, are now identical, by
allowing the reduction parameters to depend explicitly on $n_i$ (so
that they are not tensors.) This allows us to set the blocks $\cal B$
and $\cal C$ of $\cal P$ to arbitrary values for every $n_i$, and this
allows us to make arbitrary linear combinations of $v_{,A}$ and $v$
characteristic variables with arbitrary speeds, and to add arbitrary
combinations of $v_{,A}$ and $v$ to any characteristic variables made
from $w$ and $v_{,n}$, as we have discussed above. $\cal A$ being
diagonalisable is a necessary condition for $\cal P$ to be
diagonalisable, and with ${\cal B}=0$ and ${\cal C}=0$ it is also
sufficient. Therefore Definition 2b is equivalent to $\cal A$ being
diagonalisable and by Theorem 1 it is then equivalent to Definition 2a.

Alternatively, we can write down the principal part of the
second-order system in a 2+1 split as follows:
\begin{eqnarray}
\dot v &\simeq& 0, \\
\dot v_{,n}&\simeq& A_1^j (v_{,n})_{,j}+A_2 w_{,n}, \\
\dot v_{,A}&\simeq& A_1^j (v_{,j})_{,A}+A_2 w_{,A}, \\
\dot w &\simeq& B_1^{nn}(v_{,n})_{,n}+2B_1^{nA}(v_{,n})_{,A} \br
+B_1^{AB}(v_{,A})_{,B} +B_2^i w_{,i},
\end{eqnarray}
where $v_{,i}$ on the right-hand side is now considered a lower order
term. Note the way the second derivatives have been ordered
differently in $\dot v_{,n}$ and $\dot v_{,A}$. In the language of
reduction this corresponds to the reduction parameters depending
explicitly on $n_i$. The matrix $\cal P$ becomes
\begin{equation}
\partial_t
\left(
\begin{array}{c}
w \\ v_{,n} \\ v \\ v_{,A}
\end{array}
\right)
\simeq {\cal P}_{} \partial_n 
\left(
\begin{array}{c}
 w \\ v_{,n} \\ v \\ v_{,B}
\end{array}
\right) + \hbox{transverse derivatives},
\end{equation}
where 
\begin{equation} 
{\cal P}
\equiv 
\left(
\begin{array}{cccc}
B_2^n & B_1^{nn} & 0 & 0   \\
A_2 & A_1^n & 0 & 0  \\
0 & 0 & 0 & 0  \\
0 & 0 & 0 & 0 \\
\end{array}
\right).
\end{equation}
Once again $\cal P$ is diagonalisable if and only if $\cal A$
is. Either way, and taking into account the smoothness conditions, we have

\noindent{\bf Theorem 3:} {\em Definition 2b is equivalent to
Definition 2a.}


\subsection{Symmetric hyperbolicity}

Following \cite{bssn1}, we now elevate Lemma 2 to a definition of
symmetric hyperbolicity for
second-order systems in

\noindent{\bf Definition 3b:} {\em The second-order system is called
symmetric hyperbolic if and only if it admits a positive definite
energy $\epsilon$ and a flux $\phi^i$, both quadratic in $w$ and
$v_{,i}$, that obey}
\begin{equation}
\label{cons}
\dot\epsilon\simeq{\phi^i}_{,i}.
\end{equation}

Note that in obtaining (\ref{cons}) one can make arbitrary use of
$v_{,ij}=v_{,ji}$.}

We parameterise the energy $\epsilon$ by
(\ref{Hdef}), with $d_i$ replaced by $v_{,i}$, and parameterise the
flux $\phi^i$ as
\begin{equation}
\phi^i =
(w^\dagger,v^\dagger_{,m})
\left( \begin{array}{cc} F^i & F^{ik} \\ F^{\dagger im} & F^{mik} \end{array}
\right)
\left( \begin{array}{c} w \\ v_{,k} \end{array} \right) ,
\end{equation}
with $F^{\dagger i}=F^i$ and $F^{\dagger kim} = F^{mik}$. 
In the second-order system, there are no reduction parameters, and so
the non-trivial part of $HP^i$ is given by 
\begin{equation}\label{symhyp2}
\left(
\begin{array}{cc}
KB_2^i+L^iA_2 & KB_1^{ik}+L^{k}A_1^{i} \\
L^{\dagger m}B_2^i+M^{mi}A_2 & L^{\dagger m}B_1^{ik}+M^{mk}A_1^{i} 
\end{array}
\right).
\end{equation}
In a first-order system, energy conservation is precisely
equivalent to $HP^i$ being Hermitian. In the second-order system,
the relation between $HP^i$ and energy conservation is more
complicated, because $v_{,ik}=v_{,ki}$.

To see this more clearly, we write out
\begin{eqnarray}
\dot{\epsilon} &=& 2 \Bigl[
w^\dagger (KB_2^i+L^i A_2) w_{,i} \br
+ w^\dagger (K B_1^{ik} +L^{(k}A_1^{i)})v_{,ik} \br
+ v_{,m}^\dagger (L^{\dagger m}B_2^i+M^{mi}A_2) w_{i} \br 
+ v_{,m}^\dagger (L^{\dagger m}B_1^{ik}+M^{m(k}A_1^{i)})  v_{,ik}
\Bigr], \\ 
\phi^i{}_{,i} &=& 2 \Bigl[
w^\dagger F^i w_{,i} + 
w^\dagger F^{(ik)} v_{,ik} \br 
+ v_{,m}^\dagger F^{\dagger im} w_{i} 
+ v_{,m}^\dagger F^{m(ik)} v_{,ik} \Bigr].
\end{eqnarray}
Comparing the first terms in $\dot\epsilon$ and $\phi^i{}_{,i}$ we have
\begin{eqnarray}
F^i &=& K B_2^i + L^i A_2, \\
F^i&= &F^{\dagger i},
\end{eqnarray}
which admits a solution $F^i$ if and only if (\ref{T}) is
obeyed. Comparing the second and third terms we have
\begin{eqnarray}
F^{(ik)} &=& K B_1^{ik} + L^{(k}A_1^{i)}, \\
F^{\dagger im} &=& L^{\dagger m} B_2^i + M^{mi} A_2,
\end{eqnarray} 
which admits a solution $F^{ik}$ if and only if (\ref{T(ik)}) is obeyed.
Comparing the fourth terms we have
\begin{eqnarray}
\label{77}
F^{m(ik)} &=& L^{\dagger m}B_1^{ik} + M^{m(k}A_1^{i)} \equiv S^{mik},
\\
\label{78}
F^{\dagger kim}&=& F^{mik}.
\end{eqnarray}
These admit a solution $F^{mik}$ if and only if (\ref{T(mik)}) is
obeyed. It is clear that (\ref{T(mik)}) is necessary. 
To demonstrate that it is sufficient we solve explicitly for $F^{mik}$. The
general solution of (\ref{77}) is
\begin{equation}
F^{mik} = S^{mik} + J^{mik},
\end{equation}
where $J^{mik}=J^{m[ik]}$. Defining
\begin{equation}
V^{mik}\equiv S^{\dagger kim}-S^{mik},
\end{equation}
the general solution of (\ref{78}) is
\begin{eqnarray}
6 J^{mik} &=& V^{kmi}+2V^{kim}+3 V^{ikm} \br
+4V^{imk}+5V^{mik} + W^{mik},
\end{eqnarray}
if and only if $V^{(mik)}=0$, which is equivalent to (\ref{T(mik)}).
The remaing free coefficient $W^{mik}\equiv W^{[mik]}\equiv W^{\dagger
mik}$ parameterises terms in $\phi^i$ whose divergence vanishes
identically. It can be set equal to zero without loss of generality. 

We have shown that the second-order system admits a conserved energy
if and only if (\ref{T}), (\ref{T(ik)}) and (\ref{T(mik)}) hold, which
together are equivalent to (\ref{Asymm}). This is equivalent to the
existence of a conserved energy for the first-order reduction. The two
energies are in fact the same under the (unambiguous) identification of
$d_i$ with $v_{,i}$. This means that the two definitions of
symmetric hyperbolicity are equivalent, as both are equivalent to the
matrix $H$ defined by (\ref{Hdef}) being positive definite, and
obeying (\ref{Asymm}) for all $n_i$. We have

\noindent{\bf Theorem 4:} {\em Definition 3b is equivalent to
Definition 3a.}

Given an energy $H$ of the second-order system, a first-order
reduction that admits the same energy is given by the reduction
parameters (\ref{isolatev}) and $B_3^{ij}$ and ${D_j}^{ik}$ determined
in Section~\ref{section:redsymm}. Going the other way, an energy $H$
for any first-order reduction is clearly also an energy for the
original second-order system. If the second-order system admits a
multi-parameter family of energies, then some of these parameters
define reduction parameters $B_3^{ij}$ and ${D_j}^{ik}$ of the
first-order system, and the remainder parameterise the energy of that
particular reduction. An example of this split is given in
Appendix~\ref{appendix:KWB}.


\subsection{Maximally dissipative boundary conditions}

A first-order symmetric hyperbolic system on a domain $\Omega$ admits
an energy
\begin{equation}
E=\int_\Omega \epsilon \,dV
\end{equation}
whose time derivative is given by the flux through
the boundary, 
\begin{equation}
\label{84}
\dot E\simeq \int_{\partial\Omega} \phi^n \,dS,
\end{equation}
where $n_i$ is now the outward pointing normal to the boundary
$\partial\Omega$. In Appendix \ref{appendix:Hchar} we show that
symmetric hyperbolicity implies strong hyperbolicity and that the
boundary flux can be written as
\begin{equation}
\phi^n =\sum_\alpha \lambda_\alpha {\bf u}_\alpha^2.
\end{equation}
where the sum is over a suitable basis of characteristic
variables. Therefore the growth of the energy can be controlled by
controlling all characteristic variables that are ingoing at the
boundary ($\lambda_\alpha>0$) (``maximally dissipative boundary
conditions''), while ingoing or tangential characteristic variables
give negative or zero contributions to $\phi^n$ and hence $\dot E$.

The same result holds for a second-order system but, as we have seen,
the characteristic variables of the second-order system are not
unique: the $v_{,A}$ can be given arbitrary speeds, and arbitrary
combinations of $v_{,A}$ can be added to any characteristic variable.
In order to impose maximally dissipative boundary conditions with the
desired effect of controlling an energy, we need to control all
incoming characteristic variables of an actual first-order
reduction. We can do this within the second-order system by imposing
boundary conditions on characteristic variables ${\bf u}_\alpha$
constructed from $w$ and $v_{,i}$, but we need to have the correct
admixtures of $v_{,A}$ in these characteristic variables. 

We have seen that in order to show strong hyperbolicity for the
second-order system one needs to diagonalise only the matrix $\cal
A$. Let us call the characteristic variables of the second-order
system given by eigenvectors of ${\cal A}^\dagger$, and which are
therefore constructed only from by $w$ and $v_{,n}$, the ``short''
characteristic variables, and let us denote them by ${\bf
u}'_\alpha$. For proving symmetric hyperbolicity one only needs to
find a conserved energy $\epsilon(w,v_{,i})$. 

Assuming that we already have the ${\bf u}_\alpha'$ on
the one hand, and the energy $\epsilon(w,v_{,i})$ on the other, the
simplest way of generating the required ``full'' characteristic
variables ${\bf u}_\alpha$ is to make the ansatz ${\bf u}_\alpha={\rm
const}\,({\bf u}'_\alpha$ + undetermined multiples of the $d_A)$,
and to determine the coefficients and overall normalisation for
each ${\bf u}_\alpha$ so that the full characteristic variables obey
\begin{equation}
\epsilon=\sum_\alpha {\bf u}_\alpha^2.
\end{equation}
(\ref{84}) then follows.


\section{Pseudo-differential reduction method}
\label{section:pseudo}

For the purpose of comparison, we now describe the pseudo-differential
reduction method of \cite{KreissOrtiz,NOR} in our notation. We carry
out a Fourier transform in space with wave number $\omega_i$ of the
second-order system (\ref{vdot}-\ref{wdot}). We denote the Fourier
transforms of $v$ and $w$ by $\hat v$ and $\hat w$. We choose the
direction $n_i$ to be that of the wavenumber $\omega_i$ and write
$\omega_i\equiv|\omega| n_i$.  We then have $\hat d_n=i|\omega|\hat v$
and $\hat d_A=0$. We can then use $\hat d_n$ to represent $\hat
v$. The principal part of the Fourier-transformed system can be
written as
\begin{equation}
\label{pseudodifferential}
\partial_t
\left(
\begin{array}{c}
\hat w \\ \hat d_n
\end{array}
\right)
\simeq i|\omega| {\cal A}
\left(
\begin{array}{c}
\hat w \\ \hat d_n
\end{array}
\right) 
\end{equation}
where ${\cal A}$ is the matrix defined above, and the non-principal
terms not shown here are homogeneous of order zero in
$|\omega|$. In our notation, the definition of \cite{KreissOrtiz} is then

{\noindent{\bf Definition 2c:} {\em The second-order system is called strongly
hyperbolic if and only if there exists a pseudo-differential reduction
to first order of the form (\ref{pseudodifferential}) where ${\cal A}$
is diagonalisable with real eigenvalues, and the diagonalising matrix
and its inverse depend smoothly on $n_i$.}

We have set up our notation so that from Theorem 1 we immediately have

{\noindent{\bf Lemma 5:} {\em Definition 2c is equivalent to
  Definition 2a and to Definition 2b}.

As the pseudo-differential approach relies in an essential way on
Fourier transforms, it does not lend itself to defining a locally
conserved energy. Therefore there is no definition of symmetric
hyperbolicity in this approach.


\section{The evolution of constraints on the original second-order system}
\label{section:originalconstraints}

In many applications, the original second-order system is subject to
differential constraints, which are conserved under evolution. We
shall call these the ``original constraints'' to distinguish them from
the ``auxiliary constraints'' $d_i-v_{,i}=0$ that arise additionally
in the process of first-order reduction.

Note that until now we have not mentioned
or used original constraints. The reason is that in general one wants
to prove well-posedness of the second-order system {\em if the
original constraints are obeyed or not}. This is important for example
if one wants to carry out numerical simulations using ``free
evolution'' where the original constraints are imposed only on the
initial data. In the continuum the constraints then remain zero, but in
numerical free evolution they are violated through finite differencing
error. At a later time one is effectively evolving initial data that
do not obey the constraints. Therefore the continuum problem must be
well-posed for non-vanishing original constraints as a necessary
condition for numerical stability. One may of course use the original
constraints to modify the original, second-order system, but here we
assume that this has already been done, and the second-order system is fixed.

We now prove that the evolution system of original constraints is
strongly hyperbolic if it closes and if the second-order main system
is strongly hyperbolic. An equivalent result for first-order systems
subject to constraints is given in \cite{Reulastrong}.

We consider a vector $c$ of constraints which are quasilinear of the
form
\begin{equation}
c \simeq C_1^{ij} v_{,ij} + C_2^i w_{,i} ,
\end{equation}
where the matrix $C^{ij}$ is symmetric in $ij$.  It is clear from
(\ref{vdot}-\ref{wdot}) that the evolution of these constraints
is first order in space and time. If the constraint system is
closed, its principal part must then be of the form
\begin{equation}
\label{constraintevolutionprincipalpart}
\dot{c} \simeq G^i c_{,i} 
\end{equation}
for a vector of square matrices $G^i$.  Using the second-order
evolution equations (\ref{vdot}-\ref{wdot}) and comparing the 
leading order terms in $v$ and $w$ we find
\begin{eqnarray}
\label{comp1}
C_1^{(ij}A_1^{k)}+C_2^{(i}B_1^{jk)} &=& G^{(i}C_1^{jk)} , \\
\label{comp2}
C_1^{(ij)}A_2+C_2^{(i}B_2^{j)} &=& G^{(i}C_2^{j)}.
\end{eqnarray}
These identities between totally symmetric matrices hold if and only
if their contraction with $n_i$ on all indices hold for all
$n_i$. Writing the $nnn$ and $nn$ components of these equations in
matrix form we have
\begin{equation}
(C_2^n,C_1^{nn})
\left(\matrix{B_2^n & B_1^{nn} \cr A_2 & A_1^n}\right) =
G^n
(C_2^n,C_1^{nn})
\end{equation}
for all $n_i$.
We write this in compact notation as
\begin{equation}
C {\cal A}=GC.
\end{equation}
If the second-order main system is strongly hyperbolic, $\cal A$ is
diagonalisable with ${\cal A}=T\Lambda T^{-1}$.  $G$ can always be brought
into Jordan form as $G=SJS^{-1}$. Then
\begin{equation}
\tilde C\Lambda=J\tilde C, \quad \tilde C\equiv S^{-1}C T.
\end{equation}
We assume that the rows of $C$, and therefore the rows of
$\tilde C$, are linearly independent. This means that there is no
redundancy between the differential constraints, and is similar to
an assumption in \cite{Reulastrong}.

Consider now the first Jordan block of $J$ with eigenvalue
$\mu_1$. For simplicity assume it has size 2. Exceptionally writing
out the internal matrix indices, we have
\begin{eqnarray}
\tilde C_{1\alpha} \lambda_\alpha &=& 
\mu_1 \tilde C_{1\alpha} + \tilde C_{2\alpha}, \nonumber \\
\tilde C_{2\alpha} \lambda_\alpha &=& 
\mu_1 \tilde C_{2\alpha} 
\end{eqnarray}
(no sum over the index $\alpha$). From the second equation
$\tilde C_{2\alpha}=0$ for all $\alpha$ such that $\lambda_\alpha\ne
\mu_1$. Using this result and the first equation, $\tilde C_{2\alpha}=0$
precisely for those $\alpha$ for which $\lambda_\alpha=\mu_1$. By
assumption, no row of $\tilde C$ vanishes, so $\tilde C_{2\alpha}$ cannot all
vanish. Therefore, there must be at least one $\alpha$ such that
$\lambda_\alpha=\mu_1$, and the first equation must be absent, that
is, the Jordan block has only size 1. Repeating this argument for all
Jordan blocks of $J$ means that each eigenvalue of $G$ coincides with
one of $\cal A$, and that $J$ is diagonal, that is, $G$ is
diagonalisable. Writing $G=S\Lambda'S^{-1}$ where the diagonal matrix
$\Lambda'$ is a submatrix of $\Lambda$, we have
\begin{equation}
(SC){\cal A}=\Lambda'(SC).
\end{equation}
This means that the rows of $SC$ are left eigenvectors of $\cal A$,
and parameterise to characteristic variables of the second-order
system. We have shown

\noindent{\bf Theorem 5:} {\em The evolution of the original
constraints is strongly hyperbolic if the second-order main system is, and its
characteristic speeds are then a subset of those of the main
system. Furthermore, we can find a basis of characteristic
variables for the main system and the constraint system such that for
each characteristic variable ${\bf c}_\alpha$ of the constraint
system, there is a characteristic variable ${\bf u}_\alpha$ of the
main system such that}
\begin{equation}
{\bf c}_\alpha = \partial_n {\bf u}_\alpha + \hbox{ transverse derivatives}.
\end{equation}

Note that there is no such result for symmetric hyperbolicity.


\section{Symmetric hyperbolic-parabolic systems}
\label{section:symmpar}

Theorem 4.6.2 of \cite{GKO} asserts the following: Assume we have a
vector of variables $u$ and another vector of variables $z$, which
obey a linear system of evolution equations of the form
\begin{eqnarray}
\p_t u &=& D_{11} u + D_{12} z, \\
\label{D22}
\p_t z &=& D_{21} u + D_{22} z.
\end{eqnarray}
Here the $D$ are linear spatial derivative operators whose
coefficients can depend on $t$ and $x^i$. $D_{11}$ is a first-order
derivative operator such that $\p_t u=D_{11} u$ is symmetric
hyperbolic. $D_{22}$ is a second-order derivative operator such that
$\p_t z=D_{22}z$ is parabolic. $D_{12}$ and $D_{21}$ are arbitrary
first-order derivative operators. Then the coupled system is called
mixed symmetric hyperbolic/parabolic. Its Cauchy problem with periodic
boundaries is well-posed. 

The theorem can be applied straightforwardly to second-order
systems. We identify the variables $u$ of the theorem with the
variables $(v,w,d_i)$ of the first-order reduction of what is to be
the symmetric hyperbolic subsystem, and then go back to the
second-order form of this subsystem by replacing $d_i$ with $v_{,i}$. The
result is the system
\begin{eqnarray}
\label{vdot2}
\dot v&=&A_1^iv_{,i}+A_1 v+A_2 w+a +\underline{Cz}, \\
\dot w&=&B_1^{ij}v_{,ij}+B_1^iv_{,i}+B_1
v+B_2^i w_{,i}+B_2 w+b\br +\underline{D^iz_{,i}}+Dz, \\
\label{zdot}
\dot z&=& D_{22} z\br +\underline{E_1^{ij} v_{,ij}} + E_1^i v_{,i}+E_1 v
+\underline{E_2^i w_{,i}}+E_2 w. 
\end{eqnarray}
The coupling operators $D_{12}$ and $D_{21}$ are parameterised by the
matrices $C$ and $D$, and $E$, respectively. We have underlined their
principal parts to show what order of derivative is allowed in the
coupling terms.

\noindent{\bf Definition 6:} {\em A second-order system is called
  mixed symmetric hyperbolic-parabolic if it is of the form
  (\ref{vdot2}-\ref{zdot}), such that $D_{22}$ is parabolic and the
  system (\ref{vdot}-\ref{wdot}) with the same coefficients is
  symmetric hyperbolic (in the sense of Definition 2a or 2b).}

Theorem 4.6.2 of \cite{GKO} then gives us 

\noindent{\bf Lemma 6:} {\em The Cauchy problem with periodic boundary
conditions for such a system is well-posed.}


\section{Conclusions}
\label{section:conclusions}

We have formalised the definition of strong or symmetric hyperbolicity
of a system of evolution equations that are first order in time and
second order in space by reducing them to an equivalent first-order
system. We have given necessary and sufficient criteria for the
existence of a reduction that is strongly hyperbolic or symmetric
hyperbolic. These criteria are formulated entirely in terms of the
principal part of the second-order system, without an explicit
reference to the reduction. 

We have proved that the definitions of strong hyperbolicity based on a
first-order reduction, a pseudo-differential reduction, and a direct
second-order approach are all equivalent.  The definitions of
symmetric hyperbolicity based on a first-order reduction and on a
direct second-order approach are also equivalent.

In order to analyse the well-posedness of a given second-order system
in practice, there are three non-trivial calculations to be carried
out, independently of the approach in which one has defined
hyperbolicity. Suppressing technical details, they are as follows.

\paragraph*{Strong hyperbolicity} 

Strong hyperbolicity of the second-order system is equivalent to
diagonalisability, with real eigenvalues, of the matrix $\cal A$.

\paragraph*{Symmetric hyperbolicity} 

Symmetric hyperbolicity of the second-order system is equivalent to
the existence of an energy and flux quadratic in the $v_{,i}$ and $w$.

\paragraph*{Maximally dissipative boundary conditions} 

In order to impose maximally dissipative boundary conditions, one
needs the full characteristic variables of a symmetric hyperbolic
first-order reduction. This is done most easily starting from the
left eigenvectors of ${\cal A}$ and the energy
$\epsilon(w,v_{,i})$.

We have established criteria for well-posedness of the second-order
system regardless of any constraints it is subject to. However, if the
second-order system is strongly hyperbolic, and there is a closed
constraint system associated with it, then the constraint system is
also strongly hyperbolic, and the characteristic variables of the
constraint system are related to a subset of the characteristic variables
of the main system.

It is known that a first-order symmetric hyperbolic system coupled to
a parabolic system through at most first derivatives of all variables
has a well-posed Cauchy problem. We have generalised this result to
second-order symmetric hyperbolic systems through a reduction to first
order.

Appendix~\ref{appendix:KWB} gives an example of a simple second order
system discussed in both the first order reduction approach and the
direct second-order approach.


\acknowledgments

We would like to thank Fernando Barbero, Robert Beig, Gioel Calabrese,
Ian Hinder, Gabriel Nagy and Olivier Sarbach for discussions and
comments on the manuscript, and Louisiana State University, Caltech
and the Erwin Schr\"odinger Institute for hospitality.
JMM was supported by the Spanish MEC under the research
projects BFM2002-04031-C02-02 and FIS2004-01912.


\begin{appendix}


\section{Diagonalisability of matrices with a zero block}
\label{appendix:diagonalisation}

Consider the matrix
\begin{equation} \label{Mmatrix}
M=
\left( \begin{array}{cc}
A & B \\
0 & C 
\end{array} \right)
\end{equation}
where the square block $A$ has size $n$ and the square block $C$ has
size $m$.  We now prove that if $M$ is diagonalisable then $A$ and
$C$ are both diagonalisable. 

The eigenvalues of $M$ are clearly the union of those of $A$ and $C$.
The eigenvectors of $M$ can be constructed from those of $A$ and $C$
as follows: Suppose we have
\begin{eqnarray}
A v_i &=& \lambda_i v_i , \quad i=1,\ldots,n , \\
C w_j &=& \mu_j w_j , \quad j=1,\ldots,m .
\end{eqnarray}
Then we can form the eigenvectors $x_i=(v_i, 0)$, which span the
invariant subspace of $M$, and $y_j=(u_j, w_j)$ such that
\begin{eqnarray}
M x_i &=& \lambda_i x_i , \quad i=1,\ldots,n , \\
M y_j &=& \mu_j y_j , \quad j=1,\ldots,m ,
\end{eqnarray}
with the condition $(A-\mu_j)u_j = -B w_j$ for $j=1,\ldots,m$.  If
$\mu_j$ is not an eigenvalue of $A$ then $(A-\mu_j)$ can be inverted
and there is a unique solution for $u_j$. Therefore, if the
eigenvalues of $A$ and $C$ are disjoint, the eigenvectors can be
completed and the $x_i$ are linearly independent from the $y_j$
because they correspond to different eigenvalues. In this case $M$ is
diagonalisable if and only if both $A$ and $C$ are diagonalisable. 

Now suppose that $A$ and $C$ share an eigenvalue $\lambda$. 
The transformation
\begin{equation}
U M U^{-1} = \left( \begin{array}{cc}
U_1 A U_1^{-1} & D \\
0 & U_2 C U_2^{-1} \end{array}\right)
\end{equation}
with
\begin{equation}
U = \left( \begin{array}{cc} U_1 & U_1 X \\ 0 & U_2 \end{array}\right) 
\qquad
U^{-1} = 
\left( \begin{array}{cc} U_1^{-1} & -X U_2^{-1} \\ 0 & U_2^{-1}
\end{array}\right) 
\end{equation}
and $D=U_1 ( XC-AX+B ) U_2^{-1}$ brings $A$ and $C$ into Jordan form
simultaneously for suitable $U_1$ and $U_2$.  Without loss of
generality we can assume that $A$ is a single Jordan block of
eigenvalue $\lambda$ and rank$(A-\lambda I)=r$, and that $C$ is another
Jordan block with the same eigenvalue and rank$(C-\lambda I)=s$. The
matrix $M$ is diagonalisable if rank$(M-\lambda I)=0$. We have
rank$(M-\lambda I)\ge r+s$ because the $r+s$ columns of $(M-\lambda I)$
containing a 1 in the second diagonal are linearly independent,
while the matrix $B$ could provide additional linearly independent
vectors. Therefore if $M$ is diagonalisable we must have $r=s=0$ for
each Jordan block of $A$ and $C$, and so $A$ and $C$ are
diagonalisable.


\section{Hyperbolicity of the auxiliary constraint system}
\label{appendix:auxiliary}

With the choice of reduction parameters that we have used in the proof
of Theorem 1, namely (\ref{isolatev}), (\ref{mychoice1}), 
and using the following further auxiliary constraints,
\begin{eqnarray}
C_{ij} &=& c_{ij}+c_{[i,j]} = 0 , \\
C_{ijk} &=& c_{ij,k}+c_{jk,i}+c_{ki,j} = 0 ,
\end{eqnarray}
the auxiliary constraint system can be reduced to the decoupled system
\begin{eqnarray}
\dot{c}_i &=& A_1 c_i - i \mu \epsilon_i{}^{jk} c_{j,k} , \\
\dot{c}_{ij} &=& A_1 c_{ij} - i \mu \epsilon_{[i}{}^{kl} c_{kl,|j]}.
\end{eqnarray}
This is strongly hyperbolic with speeds $0,\pm\mu$ for all
$n_i$. Furthermore
\begin{eqnarray}
\dot{C}_i &=& A_1 C_i - i \mu \epsilon_i{}^{jk} C_{j,k} , \\
\dot{C}_{ijk} &=& A_1 C_{ijk},
\end{eqnarray}
which is also strongly hyperbolic (in fact, symmetric hyperbolic), and
so in this case the auxiliary constraint system is well-posed.


\section{Symmetric hyperbolicity and characteristic variables}
\label{appendix:Hchar}

Assume that $P^i$ admits a symmetriser $H$. As $H$ is Hermitian and
positive definite, there is an invertible matrix S such that
\begin{equation}
H=S^\dagger S .
\end{equation}
From this and 
\begin{equation}
\label{symm}
HP^i=P^{i\dagger}H
\end{equation}
it follows that $SP^nS^{-1}$ is Hermitian,
for any direction $n_i$. Therefore it can be diagonalised by an
orthogonal matrix $R$ (which generally depends on $n_i$), or
\begin{equation}
SP^nS^{-1}=R^{-1}\Lambda R,
\end{equation}
where $\Lambda$ is diagonal. Therefore 
\begin{equation}
P^n=T^{-1}\Lambda T, \qquad T\equiv RS, 
\end{equation}
and we have proved that symmetric hyperbolicity implies strong
hyperbolicity.

Furthermore, as $R$ is orthogonal,
\begin{equation}
H=S^\dagger(R^\dagger R)S=T^\dagger T,
\end{equation}
and so there is a preferred basis, namely the rows of $T$ (which
generally depends on $n_i$), of left eigenvectors of $P^n$ in which
$H$ is the unit matrix. In terms of the original basis
\begin{equation}
H=T^\dagger T, \qquad HP^n=T^\dagger \Lambda T.
\end{equation}
In quadratic forms the same fact can be written as
\begin{equation}
\label{sumsquares}
\epsilon=\sum_\alpha {\bf u}_\alpha^2, \qquad 
\phi^n =\sum_\alpha \lambda_\alpha {\bf u}_\alpha^2.
\end{equation}
where the sum is over the characteristic variables in the basis
encoded in the rows of $T$. 


\section{The KWB formulation of the Maxwell equations}
\label{appendix:KWB}

We use this formulation of the Maxwell equations to illustrate some of
the differences between the reduction approach and the second-order
approach, namely the existence of ``short'' and ``full''
characteristic variables, the split of the free parameters of $H$ into
those that are reduction parameters and those that are not, and the
relation between the characteristic variables of the main and
constraint systems.  We work initially in the first-order reduction
approach, and then re-interpret the same calculations in the language
of the second-order approach afterwards.

The system has been discussed in \cite{KWB,bssn1}. In flat spacetime,
in radiation gauge and in the absence of charges (source terms), it is
\begin{eqnarray}
\dot A_i &=& -E_i, \\
\dot E_i &=& - {A_{i,j}}^{,j}
+ (1-a) {A_{j,i}}^{,j} + a\Gamma_{,i} , \\
\dot \Gamma &=&(b-1) {E^i}_{,i}
\end{eqnarray}
where $v=\{A_i\}$, $w=\{E_i,\Gamma\}$ are the dynamical variables, and
$a$ and $b$ are constants that parameterise addition of the
(``original'') constraints 
\begin{eqnarray}
C_\Gamma&\equiv& \Gamma-{A_i}^{,i}=0, \\
C_E&\equiv& {E_i}^{,i}=0
\end{eqnarray}
to the evolution equations. Repeated indices have been raised with the
metric $\delta^{ij}$ and are summed over. We shall also use
$\delta_{ij}$ to decompose tensors into their normal and transverse
parts. With $d_{ij}\equiv A_{j,i}$, we consider the first-order reduction.
\begin{eqnarray}
\dot A_i &=& -E_i, \\
\dot E_i &=& - {d_{ji}}^{,j} 
+ a\Gamma_{,i} + (1-a) {{d_j}^j}_{,i} \br 
+ c\left(
{d_{ij}}^{,j} - {{d_j}^j}_{,i}\right), \\
\dot \Gamma &=&(b-1) {E^i}_{,i}, \\
\dot d_{ij} &=& - E_{j,i}.
\end{eqnarray}
The constant $c$ parameterises $B_3^{ij}$, and ${D_i}^{jk}$ has been
set to zero. For the other reduction parameters we have made the
standard choice (\ref{isolatev}), so that the $A_i$ decouple from the
$E_i$ and $\Gamma$.

The matrix ${\cal P}$, leaving out the zero rows and columns
corresponding to $A_i$, is block-diagonal with the blocks
\begin{eqnarray}
\label{scalar}
\left(
\begin{array}{cccc}
0 & a & - a & 1-a-c \\
b-1 & 0 & 0 & 0 \\
-1 & 0 & 0 & 0 \\
0 & 0 & 0 & 0 \\
\end{array}
\right)
&&
\left(
\begin{array}{c}
E_n \\ \Gamma \\ d_{nn} \\ d_{qq}
\end{array}
\right), \\
\label{vector}
\left(
\begin{array}{ccc}
0 & -1 & c \\
-1 & 0 & 0 \\
0 & 0 & 0 
\end{array}
\right)
&&
\left(
\begin{array}{c}
E_A \\ d_{nA} \\ d_{An}
\end{array}
\right), \\ 
\Bigl(0\Bigr)
&&
\Bigl(\hat d_{AB}\Bigr).
\end{eqnarray}
Here $d_{qq}\equiv q^{ij}d_{ij}$, and $\hat d_{AB}$ represents
the transverse trace-free object
$q_i^kq_j^l d_{kl}-(1/2)q_{ij}d_{qq}$. The characteristic variables
are
\begin{eqnarray}
{\bf u}_0 &\equiv& \Gamma + (b-1) d_{nn},\\
{\bf u}_\pm &\equiv& a(\Gamma-d_{nn})+(1-a-c)d_{qq}\pm \sqrt{ab} E_n, \\
{\bf u}_{\pm A} &\equiv& d_{nA}-cd_{An}\mp E_A  ,
\end{eqnarray}
with speeds $(0,\pm\sqrt{ab},\pm 1)$, and $d_{qq}$, $d_{An}$ and $\hat
d_{AB}$ with zero speed.

The first-order reduction admits the conserved energy
\begin{eqnarray}
\epsilon &=& E_i E^i + d_{ij}d^{ij}
- 2a\Gamma {d_i}^i
+(2a-1-ab) ({d_i}^i)^2 \br
+ c_1 \left[\Gamma+(b-1){d_i}^i\right]^2 
+ c [({d_i}^i)^2 - d_{ij}d^{ji}]
\end{eqnarray}
with the flux
\begin{eqnarray}
\phi^i &=& 2\left[ a(\Gamma-{d_j}^j) E^i + {d_j}^j E^i -d^{ij}E_j\right] \br 
+2c (d^{ji}E_j-{d_j}^j E^i),
\end{eqnarray}
where $c_1$ is a free parameter in the energy, and $c$ is the reduction
parameter introduced above.

We now review how one would deal with the same system in a direct
second-order approach [pointing out the relation to the first-order
approach in {\em square brackets}]. In order to show strong
hyperbolicity, one would diagonalise the matrix $\cal A$. It is
block-diagonal with the blocks
\begin{equation}
\left(
\begin{array}{ccc}
0 & a & - a \\
b-1 & 0 & 0  \\
-1 & 0 & 0  
\end{array}
\right)
\left(
\begin{array}{c}
E_n \\ \Gamma \\ A_{n,n} 
\end{array}
\right) \\
\end{equation}
and 
\begin{equation}
\left(
\begin{array}{cc}
0 & -1  \\
-1 & 0
\end{array}
\right)
\left(
\begin{array}{c}
E_A \\ A_{A,n} 
\end{array}
\right).
\end{equation}
[$\cal A$ is the sub-matrix of $\cal P$ obtained by suppressing the
rows and columns relating to $d_{Ai}$. In a different point of view,
we could set the rows and columns relating to $d_{Ai}$ in $\cal P$ to
zero by allowing the reduction parameters to depend explicitly on
$n_i$. In our example, this corresponds to setting $c=1-a$ in
(\ref{scalar}) but $c=0$ in (\ref{vector}).]

The ``short''
characteristic variables of the second-order system are obtained as
eigenvectors of ${\cal A}^\dagger$. They are
\begin{eqnarray}
{\bf u}_0' &\equiv& \Gamma + (b-1) A_{n,n},\\
{\bf u}_\pm' &\equiv& a(\Gamma-A_{n,n})\pm \sqrt{ab} E_n, \\
{\bf u}_{\pm A}' &\equiv& A_{A,n}\mp E_A  ,
\end{eqnarray}
with speeds $(0,\pm\sqrt{ab},\pm 1)$. [These are the characteristic
variables of the reduction, minus all terms in $d_{Ai}=A_{i,A}$.]

To show symmetric hyperbolicity of the second-order system, we find an
an energy. The most general one is
\begin{eqnarray}
\epsilon &=& E_i E^i + A_{i,j}A^{i,j}
- 2a\Gamma {A_i}^{,i}
+(2a-1-ab) ({A_i}^{,i})^2 \br
+ c_1 \left[\Gamma+(b-1){A_i}^{,i}\right]^2 
+ c_2 [({A_i}^{,i})^2 - A_{i,j}A^{j,i}] \br
\end{eqnarray}
with the flux
\begin{eqnarray}
\phi^i &=& 2\left[ a(\Gamma-{A_j}^{,j}) E^i + {A_j}^{,j} E^i 
-A^{j,i}E_j\right] \br 
+2c_2 (A^{i,j}E_j-{A_j}^{,j} E^i),
\end{eqnarray}
where $c_1$ and $c_2$ are free parameters. [This is identical to the energy of
the reduction, except that $c_2$ is now not a reduction
parameter. Rather, the term it multiplies is independently conserved
if we allow commutation of partial derivatives.]

In order to impose maximally dissipative boundary conditions, one
needs the full characteristic variables ${\bf u}_\alpha$. We find this
by expressing $\epsilon$ as a quadratic form in ${\bf
  u}_\alpha$. Comparing $\epsilon$ with the ${\bf u}'_\alpha$ suggests that
\begin{eqnarray}
\epsilon&=&{1\over 2ab}\left({\bf u}_+^2+{\bf u}_-^2\right)+{1\over 2}\left(
{\bf u}_{+A}{\bf u}^{+A}+{\bf u}_{-A}{\bf u}^{-A}\right)\br +\left(c_1-{a\over
b}\right){\bf u}_0^2-c_2({\bf u}_{+A}+{\bf u}_{-A})A^{n,A} \br
+\hbox{ quadratic in $A_{i,A}$},
\end{eqnarray}
with ${\bf u}_\alpha={\bf
u}'_\alpha$ + multiples of the $A_{i,A}$. [The result is
equivalent to the ${\bf u}_\alpha$ of the reduction, with $d_{ij}\to
A_{j,i}$ and $c\to c_2$.]

Finally, the constraint evolution system is
\begin{eqnarray}
\dot C_\Gamma &=& b C_E, \\
\dot C_E &=& a {C_{\Gamma,i}}^{,i}.
\end{eqnarray}
Its non-trivial characteristic variables are
\begin{equation}
{\bf c}_\pm = a \partial_n C_\Gamma \pm \sqrt{ab} C_E
=\partial_n {\bf u}_\pm + \hbox{transv. deriv.},
\end{equation}
and these expressions hold for both the second-order system and the
reduction. 

\end{appendix}


\end{document}